# Spacetime at the Planck Scale: The Quantum Computer View

Paola A. Zizzi

Dipartimento di Matematica Pura ed Applicata
Università di Padova
Via Belzoni, 7
35131 Padova, Italy
zizzi@math.unipd.it

## Abstract

We assume that space-time at the Planck scale is discrete, quantised in Planck units and "qubitsed" (each pixel of Planck area encodes one qubit), that is, quantum space-time can be viewed as a quantum computer.
Within this model, one finds that quantum space-time itself is entangled, and can quantum-evaluate Boolean functions which are the laws of Physics in their discrete and fundamental form.

# 1. Introduction

What is "space-time" at the Planck scale? Once we understand that, we will be able to formulate the theory of Quantum Gravity, the theory which should reconcile General Relativity and Quantum Mechanics. In fact, it is widely believed that at the Planck scale, the quantum aspects of gravity become relevant. Moreover, it is generally assumed that at the Planck scale, space-time is not any longer a smooth manifold, but has a discrete structure.

There are two main approaches to quantum gravity that assume quantum space-time to be discrete: Loop Quantum Gravity [1] (and spin foams [2]), and String (and M) Theory [3]. Other interesting approaches are non-commutative geometry [4], Causal Set Theory [5] and kinds of discrete models of space-time at the Planck scale, like lattice versions of loop quantum gravity [6], and Cellular Networks [7].

In our particular approach to quantum gravity, we assume discreteness of space-time at the Planck scale, and we also include the issue of information, (more precisely quantum information [8]). In fact, as it was suggested by Wheeler (the "It from bit" proposal) [9], information theory must play a relevant role in understanding the foundations of Quantum Mechanics. Wheeler's view is shared, in particular, by Zeilinger (who associates bits with elementary systems, i.e. two-level systems, and claims that the world appears quantised because information is quantised) [10].

As it was first realized by Feynmann, a quantum computer can be exponentially more powerful than a classical one in simulating a quantum system. This line of thought is what we call here the "Quantum Computer View" (QCV).

We believe that the QCV is universal, and thus can be extended to the "description" of quantum space-time itself.

Approaches similar to ours, still encompassing the QCV, are those of Lloyd [11], and Jaroszkiewicz [12].

Our approach is closely related to Loop Quantum gravity and spin networks.

Spin networks are relevant for quantum geometry. They were invented by Penrose [13] in order to approach a drastic change in the concept of space-time, going from that of a smooth manifold to that of a discrete, purely combinatorial structure. Then, spin networks were re-discovered by Rovelli and Smolin [14] in the context of Loop Quantum Gravity. Basically, spin networks are graphs embedded in 3-space, with edges labeled by spins and vertices labeled by intertwining operators. In loop quantum gravity, spin networks are eigenstates of the area and volume-operators [15].

We interpret spin networks as qubits when their edges are labelled by the spin-1/2 representation of SU (2). In this context, we use the quantum version [16] of the Holographic Principle [17].

In our model, quantum space-time is discrete, quantised in Planck units, and each pixel of Planck area, encodes a qubit. This is a quantum memory register. To process the quantum information stored in the memory, it is necessary to dispose of a network of quantum logic gates (which are unitary operators). The network must be part of quantum space-time itself, as it describes its dynamical evolution. The quantum memory plus the quantum network form a quantum computer.

In the QCV, some new features of quantum space-time emerge:

i) The dynamical evolution of quantum space-time is a reversible process, as it is described by a network of unitary operators.

ii) During a quantum computational process, quantum space-time can be in an entangled state, which leads to non-locality of space-time itself at the Planck scale (all pixels are in a non separable state, and each pixel loses its own identity).
iii) As entanglement is a particular case of superposition, quantum space-time is in a superposed state, which is reminiscent of the Many-Worlds interpretation of Quantum Mechanics [18].
iv) Due to superposition and entanglement, quantum space-time can compute a Boolean function for all inputs simultaneously (massive quantum parallelism). We argue that the functions which are quantum-evaluated by quantum space-time are the laws of Physics in their most fundamental, discrete and abstract form.
v) By scratch space management, we find that at the Planck scale it is possible to compute composed recursive functions of maximal depth.
The paper is organized as follows.
In Sect.2, we discuss the new concepts of event in quantum space-time, and its quantum information nature.
In Sect.3, we introduce the Quantum Computer View of quantum space-time at the fundamental level.
In Sect.4, we analyze the possibility of quantum space-time being in a superposed/entangled state.
In Sect.5, we investigate about a possible quantum network chosen by Nature.
In Sect.6, we illustrate how space-time can quantum-evaluate Boolean functions at the Planck scale.
In Sect. 7, we investigate about a unitary evolution of quantum space-time
Sect. 8 is devoted to the conclusions.

## 2. Qubitisation of quantum space-time

The very concept of event should be revised in the context of quantum space-time. In fact, the definition of event as a point in a four-dimensional smooth manifold becomes meaningless once space-time is assumed to be discrete, and quantized in Planck units. If the minimal length is assumed to be the Planck length: $l_P \approx 10^{-33} cm$ and the minimal time interval is assumed to be the Planck time: $t_P \approx 10^{-43} \sec$, it follows that an event in quantum space-time is an extended object without structure.
In the QCV, the quantum event encodes quantum information.
The (classical) holographic principle [17] claims that it must be possible to describe all phenomena within the bulk of a region of space of volume V by a set of degrees of freedom which reside on the boundary, and that this number should not be larger than one binary degree of freedom per Planck area.
All this can be interpreted as follows: each unit of Planck area (a pixel) is associated with a classical bit of information.
At the Planck scale, however, where quantum gravity takes place, we argue that the encoded information should be quantum, and the holographic principle should be replaced by its quantum version [16].
In the quantum version of the holographic principle, a pixel encodes one quantum bit (qubit) of information. (A qubit is a linear superposition of the logical states 0 and 1, namely: $|Q\rangle = a|0\rangle + b|1\rangle$, where a and b are complex numbers called probability amplitudes, such that $|a|^2 + |b|^2 = 1$).
The necessity of the quantum version of the holographic principle follows directly from loop quantum gravity.

In loop quantum gravity, non-perturbative techniques have led to a quantum theory of geometry in which operators corresponding to lengths, area and volume have discrete spectra.

Of particular interest are the spin network states associated with graphs embedded in 3-space with edges labelled by spins

$$j = 0, \frac{1}{2}, 1, \frac{3}{2}, ....$$

and vertices labelled by intertwining operators.

If a single edge punctures a 2-surface transversely, it contributes an area proportional to [15]:

$$l_P^2 \sqrt{j(j+1)}$$

Let us consider the edges of spin networks in the spin -1/2 representation of SU(2): they are 2-level systems, and can be thought as qubits.

In mathematical terms, the group manifold of $SU(2)$ can be parameterized by a 3-sphere with unit radius.

In fact, the most general form of $2 \times 2$ unitary matrices of unit determinant is:

$$U = \begin{pmatrix} a & b \\ -b^* & a^* \end{pmatrix} \qquad |a|^2 + |b|^2 = 1$$

where a and b are complex numbers.

For example, the action of the unitary SU(2) matrix $U_{\sigma_2} = \frac{1}{\sqrt{2}}(1 + i\sigma_2)$, where $\sigma_2$ is the Pauli matrix: $\sigma_2 = \begin{pmatrix} 0 & -i \\ i & 0 \end{pmatrix}$ on the edge states $\left|-\frac{1}{2}\right\rangle$ and $\left|+\frac{1}{2}\right\rangle$ respectively, gives the equally superposed states $\frac{1}{\sqrt{2}}\left(\left|\frac{1}{2}\right\rangle \pm \left|-\frac{1}{2}\right\rangle\right)$.

When a surface is punctured by such a superposed state, a pixel of area is created, which encodes one qubit.

The elementary pixel can then be viewed as the surface of a unit (in Planck units) sphere in three dimensions.

The pixel is punctured (simultaneously) in the poles by an edge in the superposed state of spin down and spin up.

Equivalently, a qubit corresponds to the surface of the 3-dimensional unit sphere, where the logic states 0 and 1 correspond to the poles. This is the so-called Bloch sphere.

There is clearly an analogy between the spin networks approach to quantum gravity and our Quantum Computer View of quantum space-time.

**3. Quantum space-time: is it a quantum computer?**

Having assumed that space-time at the Planck scale encodes quantum information, the latter must be processed to give rise, as an output, to the universe as we know it. If so, quantum space-time is not just a quantum memory register of n qubits: it is the whole thing, a quantum memory register plus a network of quantum logic gates.

In other words, space-time at the Planck scale must be in such a quantum state to be able to evaluate those discrete functions which are the laws of Physics in their discrete and most fundamental form. We may interpret that quantum state as the state of a quantum computer which is computing Boolean functions. But doing so, we should assume that at the Planck scale space-time is in a superposed/entangled state. In fact, any efficient quantum algorithm relies on superposition and entanglement of qubits. In quantum computation, superposition and entanglement are very important, because they allow quantum parallelism: the possibility to compute exponentially many values of a function in polynomial time.

**4. Is quantum space-time in a superposed/entangled state?**
If the qubits encoded by pixels were superposed, the surface embedding a region of space would "exist" in many different states simultaneously. This would be a quite weird wave-like aspect of quantum space-time itself. Superposition is one characteristic feature of quantum mechanics, but we should be aware of the fact that once applied to quantum space-time, it spoils the latter of its usual attributes. We think that the idea of a superposed state of qubits associated to pixels fits quite well in the Many-Worlds interpretation of Quantum Mechanics, obviously restricted to the micro-domain of space-time itself, more precisely at the fundamental level.
Do the pixels of Planck area encode qubits which are entangled to each other or not? In the affirmative, space-time itself would be spoiled of locality, at the Planck scale. In other words, two quantum events might be described by a single quantum state, each event losing its own identity. This would be a quite weird feature of quantum space-time, but it cannot be discarded *a priori*, because entanglement is a very peculiar feature of the quantum world.

Let us consider a finite number N of pixels $p_i$ (i =1, 2…N) each one encoding one qubit $|Q\rangle_i$ (notice that the number of pixels of area of a certain surface S is equal to the number of punctures made by spin network' edges in the 1/2 -representation of SU(2) onto S). The N qubits span a Hilbert space of dimension $2^N$.

The standard basis for one qubit is: $|0\rangle, |1\rangle$.

The dual basis for one qubit is:
$$\frac{1}{\sqrt{2}}(|0\rangle \pm |1\rangle)$$

The most general one qubit state is: $|Q\rangle = a|0\rangle + b|1\rangle$ where a and b are complex numbers such that:
$$|a|^2 + |b|^2 = 1$$

A 2-qubits state can be either non entangled (product state of two qubits) or entangled (a non-separable state).

The non entangled basis for two qubits is:
$$|00\rangle, |01\rangle, |10\rangle, |11\rangle$$

An example of non entangled two qubits state is the product of one dual basis vector and the qubit $|0\rangle$:
$$\frac{1}{\sqrt{2}}(|0\rangle + |1\rangle)|0\rangle = \frac{1}{\sqrt{2}}(|00\rangle + |10\rangle)$$

The entangled basis for 2-qubits (Bell states, maximally entangled) is:
$$|\Psi_\pm\rangle = \frac{1}{\sqrt{2}}(|10\rangle \pm |01\rangle)$$
$$|\Phi_\pm\rangle = \frac{1}{\sqrt{2}}(|11\rangle \pm |00\rangle)$$

## 5. The quantum network of Nature.

Let us suppose that all the N qubits encoded by N pixels are initially in state $|000.....0\rangle$. They form a quantum register of size N, but that is just storage of quantum information. To be able to perform quantum computation, the qubits of the memory must be manipulated by some unitary transformations performed by quantum logic gates (the number of the gates is called the size of the network). Now, to make a superposition of two qubits, it is necessary to dispose of the Hadamard gate:

$$H = \frac{1}{\sqrt{2}}\begin{pmatrix} 1 & 1 \\ 1 & -1 \end{pmatrix}$$

and to entangle two qubits, it is necessary the controlled -NOT (or XOR) gate:

$$XOR = \begin{pmatrix} 1 & 0 & 0 & 0 \\ 0 & 1 & 0 & 0 \\ 0 & 0 & 0 & 1 \\ 0 & 0 & 1 & 0 \end{pmatrix}$$

In the case of n qubits, we need the Walsh-Hadamard transformation:

$$H_n = H^{\otimes n}$$

Let us see how it works in the case of two qubits.
Let us write the standard basis in vector notation:

$$|0\rangle = \begin{pmatrix} 1 \\ 0 \end{pmatrix} \qquad |1\rangle = \begin{pmatrix} 0 \\ 1 \end{pmatrix}$$

The action of the Hadamard gate on the ket $|0\rangle$ is:

$$H|0\rangle = \frac{1}{\sqrt{2}}\begin{pmatrix} 1 & 1 \\ 1 & -1 \end{pmatrix}\begin{pmatrix} 1 \\ 0 \end{pmatrix} = \frac{1}{\sqrt{2}}\left(\begin{pmatrix} 1 \\ 0 \end{pmatrix} + \begin{pmatrix} 0 \\ 1 \end{pmatrix}\right) = \frac{1}{\sqrt{2}}(|0\rangle + |1\rangle)$$

and on the ket $|1\rangle$ is:

$$H|1\rangle = \frac{1}{\sqrt{2}}\begin{pmatrix} 1 & 1 \\ 1 & -1 \end{pmatrix}\begin{pmatrix} 0 \\ 1 \end{pmatrix} = \frac{1}{\sqrt{2}}\left(\begin{pmatrix} 1 \\ 0 \end{pmatrix} - \begin{pmatrix} 0 \\ 1 \end{pmatrix}\right) = \frac{1}{\sqrt{2}}(|0\rangle - |1\rangle)$$

Consider a quantum register of size two in state $|00\rangle$. The action of the Hadamard gate H on the first qubit gives the superposed state:

$$H|0\rangle = \frac{1}{\sqrt{2}}(|0\rangle + |1\rangle)$$

If we take the superposed state as the control qubit (c), and the second qubit of the memory as the target qubit (t), the action of the XOR gate is:

$$XOR : \frac{1}{\sqrt{2}}(|0\rangle + |1\rangle)_{(c)}|0\rangle_{(t)} \rightarrow \frac{1}{\sqrt{2}}(|00\rangle + |11\rangle)$$

which is an entangled state of two qubits.
A quantum memory register of size n is a collection of n qubits. Information is stored in the quantum register in binary form.
The state of n qubits is the unit vector in the $2^n$-dimensional complex Hilbert space:

$$C^2 \otimes C^2 \otimes ... \otimes C^2 \quad \text{n times.}$$

As a natural basis, we take the computational basis, consisting of $2^n$ vectors, which correspond to $2^n$ classical strings of length n:

$$|0\rangle \otimes |0\rangle \otimes ... \otimes |0\rangle \equiv |00...0\rangle$$
$$|0\rangle \otimes |0\rangle \otimes ... \otimes |1\rangle \equiv |00...1\rangle$$
$$\cdot$$
$$\cdot$$
$$\cdot$$
$$|1\rangle \otimes |1\rangle \otimes ... \otimes |1\rangle \equiv |11...1\rangle$$

In general, we will denote one basis vector of the state of n qubits as:
$$|x_1\rangle \otimes |x_2\rangle \otimes ... \otimes |x_n\rangle \equiv |x_1 x_2 ... x_n\rangle \equiv |x\rangle$$
where $x_1, x_2, ..., x_x$ is the binary representation of the integer x, a number between 0 and $2^{n-1}$.

The general state is a complex unit vector in the Hilbert space, which is a linear superposition of the basis states:
$$\sum_{x=0}^{2^n-1} c_i |x\rangle$$
where $c_i$ are the complex amplitudes of the basis states $|i\rangle$, with the condition:
$$\sum_i |c_i|^2 = 1$$

To perform computation with n qubits, we have to use quantum logic gates.
A quantum logic gate on n qubits is a $2^n \times 2^n$ unitary matrix U.
Initially, all the qubits of a quantum register are set to $|0\rangle$.
By the action of the Walsh-Hadamard transform, the n input qubits are set into an equal superposition:
$$\frac{1}{\sqrt{2}^n} \sum_{x=0}^{2^n-1} |x\rangle$$
At this point the very computation can start.

**6. Quantum function evaluation at the Planck scale**

The quantum computation of Boolean functions f is implemented by unitary operators $U_f$.

In the case of bijective functions $f: \{0,1\}^n \to \{0,1\}^n$, which are reversible, it always exists a unitary operator $U_f$ such that:
$$U_f : |x\rangle \to |f(x)\rangle.$$

where $|x\rangle$ stands (for brevity) for the input register, namely $\frac{1}{\sqrt{2}^n} \sum_{x=0}^{2^n-1} |x\rangle$.

The quantum computation of non bijective functions f: $\{0,1\}^n \to \{0,1\}^m$, (which are non reversible) requires (at least) two registers, in order to guarantee the unitary of $U_f$ (reversibility of the computation): a register of size n to keep a copy of the arguments of f, and a second register of size m, to store the values of f:
$$U_f |x\rangle |y\rangle = |x\rangle |y \oplus f(x)\rangle, \text{ where } \oplus \text{ stands for addition mod } 2^m.$$

Notice that in general, (for non trivial functions) the states $|x\rangle |y \oplus f(x)\rangle$ are entangled.

Moreover, the quantum computation of f on a superposition of different inputs, produces f(x) for all x in a single run (quantum parallelism):

$$\sum_x |x\rangle|0\rangle \rightarrow \sum_x |x\rangle|f(x)\rangle$$

But we cannot get all values of f(x) from the entangled state $\sum_x |x\rangle|f(x)\rangle$ as any measurement on the first register will yield one particular value x', and the second register will then be found with the value f(x'). It is possible, however, to compute some global properties of f(x) in a single run.

As we already said, both superposition and entanglement are necessary for quantum computation. But it is not obvious that quantum information stored in quantum space-time is exploited to perform quantum computation. It depends on which kind of quantum network (if any) Nature has chosen.

The question is: what should be computed by quantum space-time? The answer is: the global properties of Boolean functions, as in a quantum computer. In our case, we argue that the output of quantum computation would be the global structure of the Laws of Physics.

Some extra registers (called scratch space) are also needed to store intermediate results. In longer calculations (for example in computing composite functions) this leads to a large amount of "garbage" (or "junk") qubits, which are not relevant to the final result. In order not to waste space, these "junk" qubits must be re-set to $|0\rangle$ and the scratch space can then be "recycled" for further computations. Scratch space management was proposed by Bennett [19].

Let us suppose we have to calculate a composite function of depth d.

Without scratch space management, the computation would need d operations, and would consume d-1 junk registers. With scratch space management, the computation will need 2d-1 operations, and d-1 scratch registers.

For example, the computation of a composite function of depth d=2, f(x)=h(g((x))), would need 3 operations, and one scratch register, which can be reused in further computation:

$$|x,0,0\rangle \xrightarrow{U_{g12}} |x,g(x),0\rangle \xrightarrow{U_{h23}} |x,g(x),h(g(x))\rangle \xrightarrow{U_{g12}^+} |x,0,f(x)\rangle$$

Where $U_g, U_h$ are the unitary operators implementing the quantum computations of functions g and h respectively, and the suffix numbers refer to the registers operated on. The last step of the computation is just the inversion of the first step and un-computes the intermediate result. The second register can then be reused for further computations.

As we have seen, the number of required scratch registers, increases linearly with the depth of the composite function which has to be quantum computed. This fact will be very useful to our purpose.

We can imagine the boundary surface S enclosing a volume V of space, as a collection of N pixels of Planck area, each encoding a qubit. Thus S is a quantum memory register of N qubits. If all N qubits are initially set to $|0\rangle$, as always before any computation, the original register can be thought as the product of several registers: $|0\rangle_x |0\rangle_y |0\rangle_z .... |0\rangle_w$ where registers x, y, z…w have respectively size n, m, k,…, r such that n + m + k…+ r=N.

The initial quantum state $|\Psi\rangle \in C^{2^N}$ of S is then:

$$|\Psi\rangle = |0\rangle_x |0\rangle_y |0\rangle_z ....|0\rangle_w$$

Suppose that register $|0\rangle_x$ has the smallest size, for example n=2. This size is very close to the Planck scale, as for n=2, it is: $l^2 = 4l_P^2$.

The register $|0\rangle_x$ can be set to an equal superposition of basis states by the action of the Walsh-Hadamard transform $H^{\otimes 2}$ which acts locally on it:

$$H^{\otimes 2} |0\rangle_x = \frac{1}{2} \sum_{x=0}^{3} |x\rangle$$

Now the quantum state of S is:

$$|\Psi'\rangle = |x\rangle|0\rangle_y |0\rangle_z .....|0\rangle_w, \text{ where } |x\rangle \text{ stands for: } \frac{1}{2} \sum_{x=0}^{3} |x\rangle.$$

In our case, the quantum computation of a function f: $\{0,1\}^n \to \{0,1\}^{N-n}$ can be implemented by a unitary operator such that: $U_f : \sum_x |x\rangle|0\rangle_y \to \sum_x |x, f(x)\rangle$ only if the second register y has the right size to accommodate f, i.e., m=N-n, and there are no other registers available. However, if the computation of f produces n' junks bits which fill a scratch register of size n', a second register of size n', has to be provided. The best way to solve this problem, is to take a smaller first register x to enable scratch space management. Moreover, if f is a composite function f(h((g(l(….(x))))) of depth d, the original register of size N must be partitioned in such a way that there are d-1 scratch registers available. So, in order to compute highly composite functions, the first register (storing the argument) must have the smallest possible size, to leave room for the needed number of scratch registers. In particular, if n=1 (the Planck scale), the available scratch space has size N-1, and the highest level of composition for f is d=N when d-1 scratch registers, of one qubit each, sum up to the original register of size N. Thus, the quantum computation of highly composite functions must be performed close to the Planck scale, and the output (some global property of f) is obtained at macroscopic scales.

According to inflationary cosmological theories, the cosmological horizon has at present a radius $R \approx 10^{60} l_P$, thus its surface area is $A \approx 10^{120} l_P^2$, that is an area of $10^{120}$ pixels, each one encoding one qubit. In the QCV, the cosmological horizon's surface can be interpreted as a quantum memory register of $N = 10^{120}$ qubits. Thus, space-time at the Planck scale can compute a composite function of maximal depth $d = 10^{120}$.

Cosmological models based on the QCV have been considered by the author [20] and by Lloyd [21].

As we have already said, we believe that the recursive functions computed by quantum space-time at the Planck scale are the laws of Physics in their discrete, abstract, and fundamental form.

## 7. Unitary evolution and its consequences

The quantum evolution of a quantum computer is described by unitary operators, and this guarantees a reversible computation.

It follows that, in the quantum computer view, the dynamical evolution of quantum space-time itself is a reversible process. This sounds like a paradox, as far as we think of quantum space-time as a pre-space-time with almost all the same characteristics of classical space-time, which is the seat of irreversibility.

Irreversibility might be just an emergent feature at larger scales.

One should be able, however, to figure out what it means reversibility of quantum space-time itself. The simplest answer leads us back to Wheeler's "space-time foam" [22], made up of virtual black holes (and wormholes). Like all virtual processes, also this one takes place by virtue of the time-energy uncertainty relation, (which at the Planck scale is saturated). A quantum black hole of Planck mass, comes into existence out of the vacuum, and then evaporates in Planck time, releasing a quantum of Planck energy back to the vacuum. As this "virtual" process is due to quantum fluctuations of the vacuum, which are non-dissipative [23], it can be considered a reversible process, unless a measurement takes place. But virtual particles cannot be probed.

## 8. Conclusions

The QCV of space-time at the Planck scale relies on linear concepts like superposition and entanglement. Thus, this view cannot be extended to the macroscopic domain, where space-time is described by the non linear equations of General Relativity. To understand how, from the linearity of the Planck scale level we obtain the non linearity of the classical macroscopic level, it might be useful to consider self-organizing models and related technicalities. This is what we call emergence of classicality and complexity (our classical world emerges as one which is complex).

As we have seen, in the QCV, quantum space-time looks like having a reversible dynamical evolution. But what does it mean that space-(time) evolves in time, and moreover in a reversible manner? As we have seen, this paradox can be solved by assuming Wheeler's picture of "space-time foam" which however excludes time flow at the Planck scale.

Thus, both non linearity and irreversibility, which have no home in the QCV, should be emergent features of space-time.

In the QCV, also locality is lost: "space-time" itself is non local at the Planck scale, due to the entanglement of pixels/qubits. This is very much on line with Penrose's argument, stating that the theory emergent from spin networks should have a fundamentally non-local character [24].

As far as causality is concerned, it is a more subtle point. However we believe that, because of non-locality due to entanglement of pixels, micro-causality is missing at the Planck scale, at least in its usual form.

Finally, despite all these weird features, space-time at the Planck scale seems to be able to compute its own dynamical evolution, by quantum evaluating recursive functions.

## Acknowledgements
I wish to thank G. Peruzzi e G. Sambin for useful discussions.